\documentclass[ superscriptaddress, twocolumn, secnumarabic, tightenlines, nobibnotes, aps]{revtex4}
\usepackage{bm}
\usepackage{amsmath,amssymb,amsfonts,epsfig,graphicx}

\usepackage{colordvi}

\begin{document}

\title{The high temperature expansion of the classical  $XYZ$  chain}

\date{\today}

\author{E.V. Corr\^ea Silva}
\affiliation{Departamento de Matem\'atica e Computa\c c\~ao,
Faculdade de Tecnologia, 
Universidade do Estado do Rio de Janeiro.
Estrada Resende-Riachuelo, s/n$^{\textit o}$,
Morada da Colina, CEP 27523-000,  Resende-RJ, Brazil}

\author{Onofre Rojas}
\affiliation{Departamento de Ci\^encias Exatas, Universidade Federal de
Lavras, Caixa Postal 3037, CEP 37200-000, Lavras-MG,  Brazil}

\author{James E.F. Skea}
\affiliation{Instituto de  F\'{\i}sica, Universidade do Estado do Rio de Janeiro,
R. S\~ao Francisco Xavier n$^{\textit o}$ 524, Bloco B,
CEP- 20559-900, Rio de Janeiro -RJ, Brazil}

\author{S.M. de Souza}
\affiliation{Departamento de Ci\^encias Exatas, Universidade Federal de
Lavras, Caixa Postal 37, CEP 37200-000, Lavras-MG,  Brazil}

\author{M.T. Thomaz}
\email[Corresponding author: ]{mtt@if.uff.br}
\affiliation{Instituto de F\'{\i}sica,
Universidade Federal Fluminense, Av. Gal. Milton Tavares de
Souza s/n$^{\textit o}$, CEP 24210-340, Niter\'oi-RJ, Brazil}

\begin{abstract}

We present the $\beta$-expansion of the Helmholtz 
free energy of the  classical $XYZ$ model, with a 
single-ion anisotropy term  and in the presence of an external 
magnetic field, up to order $\beta^{12}$.  We compare our 
results to the numerical solution of  Joyce's 
[Phys. Rev. Lett. {\bf 19}, 581 (1967)]  expression  for the thermodynamics
of the $XXZ$ classical model,   with neither single-ion 
anisotropy term  nor external  magnetic field. This 
comparison shows that the derived analytical expansion is valid for
intermediate temperatures such as $kT/J_x \approx 0.5$. 
We show that the specific heat and magnetic susceptibility of the
spin-2 antiferromagnetic chain can be approximated by their 
respective classical  results, up to $kT/J \approx 0.8$,
within  an error of $2.5\%$. In the absence of an external
magnetic field, the ferromagnetic and antiferromagnetic chains
have the same classical Helmholtz free energy. We show how this two types
of media react to the presence of an external magnetic field.

\end{abstract}

\maketitle

\section{Introduction}

The classical limit of quantum spin chains has attracted 
attention since  this class of models was  proposed to 
describe the magnetic interactions
of one-dimensional models. In 1964 Fisher\cite{fisher} calculated 
the analytical expression of the Helmholtz free energy (HFE) of the
$XXX$ chain in the presence of an external 
magnetic field.  Some years later, Joyce\cite{joyce} included in that model an
anisotropy term of interaction between the $z$ components of 
neighbouring spins, and obtained its
exact HFE for a vanishing external magnetic field.
Since then  the thermodynamics of the classical $XXX$ chain, in the presence of
in-plane or transverse magnetic field has been studied at very low
temperatures (limit of long wavelength) and the mapping of this
 classical  spin chain into the Sine-Gordon
model\cite{sine-gordon} has  permitted one  to
 obtain analytical expressions for some 
termodynamical functions in this region of temperature. 
On the other hand,  the classical
$XXZ$ chain in the presence of an external magnetic field in the
$z$ direction  has also been studied numerically\cite{delica}.

Recently we obtained the high temperature expansion (HTE)  of the
HFE of the quantum spin-$S$ XYZ
chain\cite{brief}, with a single-ion anisotropy term and in the 
presence of an external magnetic field, up to order 
$\beta^5$. (Here we have $\beta= \frac{1}{kT}$, where 
 $k$ is the Boltzmann's constant, $T$ is the temperature in Kelvin.)
The HFE of the classical model is obtained from Ref.  \cite{brief} by
taking the limit $S\rightarrow\infty$, where $S$ is the spin value
($S = 1/2, 1, 3/2, 2, \cdots, \infty$).  The calculation of this 
thermodynamical function of the quantum model was done by using
the cummulant series for the one-dimensional models presented in
Ref.  \cite{chain_m}. This approach can also be 
applied to classical periodic chain  models with
 nearest-neighbor interactions.
To the best of our knowledge,  analytical expressions of the
 HFE  for intermediate and high temperature regions
of the classical XYZ  model are currently unknown.

In Ref.  \cite{brief},  we showed that certain thermodynamical
functions like the magnetization and magnetic susceptibility per 
site of the quantum $XYZ$ chain can be well approximated,
 in the intermediate and high temperature regions, 
by their classical results for $S \geq 3/2$. The absence of results
for the quantum $XYZ$ chain for $S>1/2$, makes the knowledge
of  the classical model  very interesting in the region of $kT\gtrsim  J_x$,
where  $J_x$ is the coupling constant between the closest spins  along the 
$x$ direction.  Even for other  thermodynamical functions, like
the specific heat per site, the classical models
give results with the correct order  of magnitude 
for those regions of temperature. Nowadays  the possibility 
of designing  different materials with suitable properties 
turn  the analytical expressions of the classical
thermodynamical functions  to be very helpful.

In Ref.  \cite{brief} we calculated the 
$\beta$-expansion of  the HFE of the quantum and 
classical $XYZ$ model, with single-ion anisotropy term and in 
the presence of an external magnetic field, up to order $\beta^5$. 
When applying  the method of Ref.   \cite{chain_m} directly to
the classical model we have the advantage of calculating 
traces of   $c$-numbers, thus diminishing the number of
terms to be evaluated.  Here  is the first time that this method
 is being applied directly  to a classical   periodic chain 
with first-neighbour interactions.

In section 2,  we present the Hamiltonian that describes the 
classical $XYZ$ chain with unitary spin ($|\vec{s}_i |=1$)
at  the $i$-th site.  In  this   model we take 
into account the presence of a single-ion anisotropy term and
an external magnetic field along the $z$ direction. In section 3,  we 
present the results of Ref.\cite{chain_m} when applied directly
to the classical $XYZ$    chain;  a rule is obtained that 
allows, in the present case,   optimization of the algebraic
 calculations of the HFE.  In section 4,  we compare 
Joyce's solution\cite{joyce} with  our results for the
 particular case of   a  classical $XXZ$ chain 
($J_x = J_y$) without the single-ion anisotropy term ($D=0$) and in the
absence of an external magnetic field ($h=0$),  
for intermediate and high temperatures.
 Joyce's  solution  of the classical $XXZ$ chain 
allows one to obtain  the  behaviour  of the specific 
heat per site at low  temperatures, and then apply Pad\'e's method
to extend the validity of the HTE to  lower temperatures. In section 5
we study some thermodynamical functions of the classical $XYZ$ chain
in the presence of an external magnetic field. Finally, in section 6
we present a summary of our results.

In appendix \ref{Apend_A}, we present some useful integral results.
As an example of our HTE of the HFE of the
classical $XYZ$ chain, we present in appendix \ref{Apend_B} its expansion 
up to order $\beta^6$ in the absence of an external magnetic 
field. Its complete expansion, up to order $\beta^{12}$, 
is  quite lengthy and can be obtained under request to the autors.


\section{ The classical $XYZ$ chain  with unitary spin}\label{hamiltonian}

The classical version of the 
hamiltonian  of the anisotropic  $XYZ$  chain  with
 unitary spin-$S$, as shown in Eq.(2) of  Ref.  \cite{brief},  is

\begin{eqnarray}\label{1}
{\mathbb H} = \sum_{i=1}^{N} {\textbf H_{i,i+1}} &=& 
 \sum_{i=1}^{N} \{ J_x s^{x}_{i}s^{x}_{i+1}+J_y s^{y}_{i}s^{y}_{i+1}+
J_z s^{z}_{i}s^{z}_{i+1}  \nonumber \\
%
%
 & & \hspace{1cm} - h s^{z}_{i} + D (s^{z}_{i})^2 \} ,
\end{eqnarray}

\noindent where
\vspace{-0.5cm}

\begin{eqnarray}  \label{2}
s_i^{x} & =&   \sin(\theta_i) \, \cos(\phi_i), \hspace{0.2cm}
s_i^{y}  =    \sin(\theta_i)  \, \sin(\phi_i) ,  \nonumber \\
%
%
s_i^{z} & =& \cos(\theta_i); 
\end{eqnarray}

\noindent  $\theta_i$ and $\phi_i$ are  the 
polar and azimuthal angles, respectively,  of the classical spin $\vec{s}_i$; 
$N$ is the number of sites in the periodic chain;  $h$ is the 
external magnetic field along the $z$-axis and $D$ is the single-ion 
anisotropy parameter. The constants $J_x$, $J_y$ and $J_z$
give the strength of first-neighbour interactions between the 
components of the spins.  From (\ref{1}) and (\ref{2}), we define the function 
${\cal H}$, so that

\begin{equation}\label{h_angles}
{\mathbb H} = 
\sum_{i=1}^{N} {\cal H}(\theta_i, \phi_i, \theta_{i+1}, \phi_{i+1}) .
\end{equation}


\section{The  method of cummulant series applied to classical chains}\label{method}

In Ref.  \cite{prb67_03} we  presented a  survey of 
the cummulant series method and its application to  one-dimensional 
periodic {\em quantum} chain models with first-neighbour interactions.
Here, we discuss the application of this method to any {\em
classical} one-dimensional chain model subject to periodic boundary
conditions, spatial translation invariance, and nearest-neighbor
interactions. Following Ref. \cite{chain_m}, we obtain the analytical
expressions for the HTE of the HFE in the thermodynamical limit of
such models,  which can   be written as the $\beta$-expansion

\begin{eqnarray}
{\mathcal{W}(\beta)} = 
- \frac{1}{\beta} \left[\ln{({\rm tr}_i(\bf{1}_i))} + \ln{(1 + \xi(\beta))}
\right],       \label{3}
\end{eqnarray}

\noindent where ${\rm tr}_{i}({\bf 1}_{i})$ equals
the dimension of the classical Hilbert space of the $i$-th site, 

\begin{eqnarray}
\xi (\beta) = \sum_{n=0}^{\infty} \frac{1}{(n+1)!} \left.
\frac{\partial{^n}}{\partial\lambda^n}(\varphi(\lambda)^{n+1})
 \right|_{\lambda=1}
                  \label{4}
\end{eqnarray}

\noindent and the auxiliary function $\varphi(\beta)$ is given by

\begin{eqnarray}
\varphi(\lambda) =  \sum_{m=1}^{\infty} \;\;  \sum_{n=m}^{\infty}
  \frac{(-\beta)^n}{\lambda^m} \,  H_{1,m}^{(n)} .
          \label{5}
\end{eqnarray}

\noindent The functions $H_{1,m}^{(n)}$ correspond to the
``connected'' strings with $n$ operators ${\textbf H}_{i, i+1}$
(in the present case, see Eq. (\ref{1})) so that $m$ of 
them are distinct, that is,

\begin{eqnarray}
H_{1,m}^{(n)}=
{\sum_{\{n_i\}}^{n}}{^{\prime\prime}} \big\langle
        \prod_{i=1}^{m}\frac{{\textbf H}_{i,i+1}^{n_i}}{n_i!}\big\rangle.
               \label{6}
\end{eqnarray}

The normalized trace is defined as

\begin{eqnarray}
& &\langle {\bf H}_{i_1, i_1 + 1} \ldots {\bf H}_{i_m, i_m +1} \rangle
 \equiv   \nonumber \\
%
%
&& \hspace{0.5cm} 
 \frac{{\rm tr}_{i_1, \ldots,  i_m+1}  ( {\bf H}_{i_1, i_1 + 1} 
 \ldots {\bf H}_{i_m, i_m +1}) }
{ {\rm tr}_{i_1}( {\bf 1}_{i_1})  
    \ldots  {\rm tr}_{i_m + 1}( {\bf 1}_{i_m +1})},
\label{6.1}
\end{eqnarray}

\noindent where the indices $i_1, \ldots, i_m$ can   be equal or distinct.
The notation ${\rm tr}_{i_1, \ldots, i_m +1}$ stands for the
trace over the degrees  of freedom of   $m +1$ distinct sites in the set:
$\{i_1, \ldots, i_m +1\}$, with $ m \leq n$. In addition,  
${\rm tr}_{i_k}$ represents the trace over the degrees of freedom of the
$i_k$-th site and ${\bf 1}_{i_k}$ is its identity operator.

Eq.(\ref{6}) differs from its quantum version on showing a
normalized trace, rather then a $g$-trace: in the classical case, all
${\textbf H_{i,i+1}}$ are commuting functions. This property greatly
simplifies all calculations (see Appendix \ref{Apend_A}).
The notation
{\scriptsize$\underset{\{n_i\}}{\overset{n}{\sum}}{^{\prime\prime}}$}
stands for the restriction
$\underset{i=1}{\overset{m}{\sum}}{^{\prime\prime}} \; n_i=n$ and
$n_i\ne 0$ for $i=1,2,..,m$. The index $m$ satisfies the condition
$1\leqslant m \leqslant n$. 
Equations (\ref{3}) to (\ref{6})  are valid
for any  classical one-dimensional chain model
subject to periodic boundary conditions,
spatial translation invariance, and nearest-neighbors interactions. 
In order to reach   higher orders in  $\beta$ in the HTE
of the HFE, it is important to optimize the calculation of the normalized 
traces  in  Eq. (\ref{6}). 
Although  the classical model has less terms to
 be calculated than  its quantum version,
we still have a large number of integrals to be done 
for each  order in $\beta$. For Hamiltonian (\ref{1}) it is 
straightforward   to verify 
that, at each site, using the results 
(\ref{A1}) and (\ref{A2}) of appendix \ref{Apend_A},
that we only have  a non-vanishing normalized trace
$\langle (s^x)^l  (s^y)^m (s^z)^p \rangle$, if and only if
$l$, $m$ and $p$ are  all even.

\section{The $\beta$-expansion of the HFE of the classical  
$XYZ$ chain}\label{expansion}

>From the results of Ref. \cite{brief}, we can obtain the 
HTE of the HFE of  the classical $XYZ$ model, up to order 
$\beta^5$, by taking the  limit of $S \rightarrow \infty$ in its Eq.(3). 
By using the results of section \ref{method}, we obtain here the HTE of
the HFE, ${\mathcal W}_{class}$, up to order $\beta^{12}$, for a
non-vanishing external magnetic field $h$ and in the
presence of a single-ion anisotropy term  in the Hamiltonian
(\ref{1}).  The derived expansion is very lengthy and it can be
 obtained from the authors by request.
Although this series has a large number of terms, it 
can be easily handled  by any  CAS system (we used {\tt Maple}). 
In appendix \ref{Apend_B} we present the expansion of the HFE of the
classical $XYZ$ chain, with unitary spin (see Hamiltonian (\ref{1}),
in the absence of an external magnetic field ($h=0$)), up to order
$\beta^6$. The resulting expansion (\ref{B1}) agrees with the series
of this thermodynamical function obtained from eq.(3) of Ref.
\cite{brief}, in the limit $S\rightarrow \infty$.

In this section,  we have
$J_x = J_y = J$ and $J_z = J \Delta$, for the constants
 in  Hamiltonian (\ref{1}) for the particular case 
of the classical $XXZ$   model.

\subsection{Comparison of the HTE of the  classical $XXZ$ chain with Joyce's 
solution}

Next, we perform some comparisons with known results, 
to verify the interval of temperature where our HTEs still hold.
Joyce\cite{joyce}  obtained the exact solution of  the
classical $XXZ$ chain,  without  a single-ion anisotropy
 term ($D=0$) and in the  absence of an external magnetic field 
($h=0$). In order to check  our HTE of  the
classical model and  to verify how it extends our previous
results\cite{brief},  we compare in Figs. \ref{fig_1}a 
and  \ref{fig_1}b   the  Joyce's numerical solution for the HFE
with our expansions  up to orders $\beta^5$ and $\beta^{12}$.
In Figs. \ref{fig_1}c and \ref{fig_1} d we present
the respective  percental difference between each 
of those expansions and the
exact Joyce's solution.  In Fig. \ref{fig_1}a   we plot the HFE 
for  $\Delta = \pm 0.5$. At  first glance of this graph suggests
that the   expansion up to $\beta^{12}$ is poorer than the 
expansion up to $\beta^5$.  However, a closer look at the interval 
$kT/J = [ 0.4, 1 ]$ (see the detail box) shows
that the $\beta^{12}$-curve coincides with the exact solution up to 
$kT/J = 0.53$, whereas the $\beta^5$-curve
coincides with it only up to $kT/J = 0.9$. 
In Fig. \ref{fig_1}b, we take $\Delta = \pm 1$ and we 
see that the  expansion  up to $\beta^5$ coincides with the numerical
solution of Joyce's expression up to $kT/J = 0.56$
whereas  the series up to $\beta^{12}$  goes up to 
$kT/J = 0.39$, which can certainly be considered an intermediate 
region of temperature.

In  Fig. \ref{fig_2} we compare the specific heat per 
site,  
$C_{class} (\beta) = - \beta^2  
   \frac{\partial ^2 }{\partial \beta^2} \left(\beta {\mathcal  W}_{class}(\beta)\right)$,
 obtained from Joyce's solution of the classical $XXZ$ 
 chain (with $h=0$  and $D=0$) and the HTE up
to $\beta^6$ (see Ref. \cite{brief}) and up to $\beta^{13}$. We also include
(Figs. \ref{2}c and \ref{2}d) the respective relative 
percental  difference of these expansions to the exact solution. In Fig.
\ref{fig_2}a we have $J=1$ and $\Delta = \pm 0.5$, whereas in Fig.
\ref{fig_2}b we have $J= 0.5$ and $\Delta = \pm 2$. Again, in both plots
we verify that the $\beta^{13}$ expansion extends to lower temperature
the validity of the HTE of the classical model in Refs. \cite{brief,ejpB}, 
as shown in the detail box of Fig. \ref{fig_2}c.
It is simple to derive from   Joyce's solution
the correlation function between  the $x$, $y$ and $z$  components 
of the classical spins  of  first nearest neighbours.   
Due to the symmetry in 
the $x$ and $y$ directions in Hamiltonian (\ref{1}), 
 for $J_x = J_y$, we have 
$\langle s_i^x   s_{i+1}^x \rangle = \langle s_i^y   s_{i+1}^y \rangle$. 
In Figs. \ref{fig_3} we plot the correlation function 
$\langle s_i^x   s_{i+1}^x \rangle 
         = \frac{ \partial {\mathcal  W}_{class}}{\partial J_x}$
for $J=1$ and $\Delta = 0.5$
(Fig. \ref{fig_3}a)  and for   $J=0.5$ and $\Delta = 2$ (Fig. \ref{fig_3}b),
derived from the exact result and from our $\beta$-expansion 
(up to order $\beta^{12}$)  for  the HFE of the classical 
model. From Fig. \ref{fig_3}b  we see that  the $\beta^{12}$-curve 
 gives the correct maximum of 
the function $\langle s_i^x   s_{i+1}^x \rangle$ for $J=0.5$ and 
$\Delta= 2$. 

In Figs. \ref{fig_4} we plot the correlation function 
$\langle s_i^z   s_{i+1}^z \rangle 
= \frac{ \partial {\mathcal  W}_{class}}{\partial J_z}$
for the exact result  of the classical model and our 
$\beta$-expansion,   for the same set of
constants $J$ and $\Delta$ as in Figs. \ref{fig_3}. From 
Fig. \ref{fig_4}a  we see that the derived  HTE up to $\beta^{12}$
gives the correct maximum of this thermodynamical  function.

One way to extrapolate the results of our high temperature  
series to higher orders in $\beta$ is through  the Pad\'e  approximants 
(PA), which allows us to combine thermodynamical information from
 both high and low temperatures.  Among the several approaches
 to Pad\'e approximants\cite{outros},  here we employ the two-point
 Pad\'e approximant\cite{pr_antigo} to extend our  $\beta$-expansion 
 of  the specific heat per site to low temperatures.

>From the numerical analysis of Joyce's solution\cite{joyce} of the
specific heat per site at very low temperatures, for $ J\ne 1$ and
$\Delta \ne 1$, we realize that it has a polynomial behaviour in $T$.
For the range of temperature $T\in[0, 0.1]$, it can be chosen
to be of fourth order in $T$, $C (T)\approx
1+a_1T+a_2T^2+ a_3T^3+ a_4 T^4$, where the coefficients $a_i$ can be
adjusted appropriately by linear regression. In this case, we have 5
known terms in the region of low temperature and 12 known terms in the
region of high temperature, resulting in a Pad\'e approximant with 17
terms.

Fig.\ref{fig_5}a compares results for the specific heat per site,
namely the best PAs and Joyce's exact solution, for $J=1$ and $\Delta
= 0.5$.  For this constants,
we have: $a_1=0.51519$, $a_2=1.42213$, $a_3 =-2.10687$ and
$a_4=62.1676$.  Below the graph in  \ref{fig_5}a, we show the 
percental  deviation of each PA with respect to the numerical
 result of the exact  solution.  It turns out that 
 P$_{11,6}$    is the   best approximation of the 
   exact result, within a difference 
of less than $2\%$   for all values of  temperature.

Fig. \ref{fig_5}b shows a similar comparison, but for the parameters
values   $J= 0.5$  and $\Delta = 2$. For those constants, 
we obtain numerically:  $a_1=0.529029$, 
$a_2=1.63073$, $a_3=-8.64101$ and $a_4=86.746$.  In 
Fig.  \ref{fig_5}c,  we present the  percental deviation of 
each PA to the numerical result of the exact solution.  
We see  that  P$_{10,7}$  is the best approximation
 to  the exact  result. In the whole
interval of temperature,   its   percental  difference to the exact 
solution is less than  $10\%$ . Although Figs. \ref{fig_5}a and 
\ref{fig_5}b refer to the same   thermodynamical function, each case 
demands specific PAs  for the best fitting.

\subsection{The  classical model as an approximation to the quantum model.}

The classical $XYZ$ chain  is a good approximation to  the quantum 
chain,  for all values of $S$ ($S= 1/2, 1, 3/2, 2, \cdots$), in the high 
temperature region ($J_x  \beta \ll 1$). In Ref. \cite{ejpB} we showed
that some thermodynamical functions of the quantum  $XXZ$ chain
can be well approximated by their  classical version for 
not so high temperatures.  This region of temperature, where the classical
and quantum models are equivalent,  depends on the termodynamical function 
and  on the   spin.  The higher the spin of the quantum model, the more involved its
numerical solution gets, due to the growing number degrees of freedom.
However, as far as the $XYX$ model is concerned, for $S\geq2$, the
quantum HTE of a given thermodynamical quantity can thus be well
approximated to its classical HTE.

Some materials are well described by $XXZ$ models
with higher values of spin. For example, the 
$({\rm C}_{10} {\rm H}_8{\rm N}_2){\rm MnCl}_3$,
described by a $S=2$ model\cite{granroth};
and  the $({\rm CH}_3)_4{\rm NMnCl}_3$,
also known as TMMC, described by a $S=5/2$ model\cite{birgeneau,hutchings}.
In Ref. \cite{yamamoto96},   Yamamoto carried out  Monte  
Carlo calculations  of the  thermodynamics of the $S=2$ $XXZ$
chain with 96 sites,  with $\Delta=1$, $D=0$ and $h=0$. He obtained
the temperature dependence of the specific heat per site 
and the magnetic susceptibility per site
for any temperature.   Fig. \ref{fig_6}a compares
the numerical results of Yamamoto \cite{yamamoto96} for the
specific heat per site,   for the $S=2$ antiferromagnetic case, to
our $\beta$-expansion of the   {\em classical} specific   heat
per site, up to order $\beta^{13}$.   Fig. \ref{fig_6}b shows the relative 
percental error between this two curves,  which is less than $2.5 \%$ up to $kT/J=0.75$.
In Fig. \ref{fig_7}a we compare 
Yamamoto's quantum  magnetic susceptibility per site 
$\left( \chi = - \frac{\partial^2 {\mathcal W}}{\partial h^2} \right)$
of the $S=2$  antiferromagnetic  chain\cite{yamamoto96}
to its  equivalent classical function 
obtained from our $\beta$-expansion, up to order $\beta^ {12}$. 
In the same token, Fig. \ref{fig_7}b shows their percental
difference, which is smaller than $2.5\%$ up to
 $kT/J \sim 0.8$.

In Ref. \cite{ejpB} we verified that the higher the spin, 
the closer quantum and classical thermodynamical functions 
get. The curves, for a given thermodynamical 
 function of the quantum models,  do not cross  for different values of spin.  
Although  we do not have a  numerical study of the
 thermodynamics of the antiferromagnetic 
chain with $S=5/2$, we can  affirm that its specific heat
and magnetic susceptibility  per site, 
for $kT/J$ up to $0.8$,  differs from the classical result 
in  less than $2.5\%$.  This also applies to the
TMMC\cite{jensen},  since it anisotropy in the $z$ directions 
is small, namely  $\Delta= 0.016$.

\section{The thermodynamical functions of the classical $XYZ$ chain}

>From expression (\ref{B1}) of Appendix \ref{Apend_B}, we verify that the HFE of 
the classical $XYZ$ chain  in the absence of an external magnetic
 field ($h=0$),  is an even function of   the constants $J_x$,  $J_y$ and $J_z$, 
so that this function  is the same for classical ferromagnetic and antiferromagnetic 
materials  with the same single-ion anisotropy $D$-term. As a consequence of this
property, we have that some thermodynamical functions
have well defined parity under the transformation 
$(J_x,  J_y, J_z) \rightarrow ( - J_x,  - J_y,  - J_z)$:
the specific heat,  $\langle (s_i^z)^2 \rangle$, the entropy 
 and the mean energy are even functions, whereas the 
 first-neighbor correlation functions,
$\langle s_i^x s_{i+1}^x \rangle$,
$\langle s_i^y s_{i+1}^y \rangle$   and
$\langle s_i^z s_{i+1}^z \rangle$ are odd functions.
None of the quantum versions of the above mentioned functions
have defined parity for $h=0$ \cite{brief}.

Figs. \ref{fig_8} show how each type of medium responds 
to the presence of an external magnetic
field. We use the  fact  that the mean energy 
per site,   $\langle {\mathcal E}\rangle = 
    \frac{\partial(\beta {\mathcal W}_{class})}{\partial \beta}$,  
is the same for both ferromagnetic and antiferromagnetic 
media at $h=0$.    Fig. \ref{fig_8}a shows
the curve of the mean  energy as  a function of $J_x \beta$,
for $J_y/J_x = 1/3$, $J_z/J_x= \pm 2/3$  and $D/J_x= -0.6$, 
at $h=0$. 

Fig. \ref{fig_8}b shows, for the ferromagnetic ($J_z/J_x = -2/3$) and
antiferromagnetic ($J_z/J_x = 2/3$) cases, the difference of the mean
energy per site with and without external magnetic field
(cf. Fig. \ref{fig_8}b, 
$\Delta {\mathcal E} = \langle {\mathcal E}\rangle|_{h} 
- \langle {\mathcal E}\rangle|_{h=0} $), 
plotted as a function of $h/J_x$, for $J_x \beta=0.5$
and $J_x \beta= 1$. Fig. \ref{fig_8}c quantifies this difference,
showing the corresponding percental differences
($\Delta {\mathcal E} (\%) = \frac{\langle {\mathcal E}\rangle|_{h} 
- \langle {\mathcal E}\rangle|_{h=0}}{ \langle {\mathcal E}\rangle|_{h=0}}
\times 100\% $).

The behavior of the mean square of the $z$-component of each spin in
the chain at $h=0$ as a function of $J_x \beta$, $\langle (s_i^z)^2\rangle
= \frac{\partial {\mathcal W}_{class}}{\partial D}$, can be seen in
Fig. \ref{fig_9}a. The reaction of the ferromagnetic and
antiferromagnetic media to the presence of an external magnetic field,
$\langle (s_i^z)^2\rangle - \langle (s_i^z)^2\rangle|_{ h=0}$,
can be seen in Fig. \ref{fig_9}b, as a function of $h/J_x$
at $J_x \beta= 0.5$ and $J_x \beta =1$.
For both graphs, we take the same set of values for $J_y/J_x$,
$J_z/J_x$ and $D/J_x$ as  that of
Fig. \ref{fig_8} for the ferromagnetic and antiferromagnetic media. 
Fig. \ref{fig_9}c is similar to Fig. \ref{fig_9}b, showing 
the corresponding percental difference with 
respect to $\langle (s_i^z)^2\rangle|_{ h=0}$.

In Fig. \ref{fig_10} we compare the magnetization per site, 
$ M_{class} = -\frac{\partial {\mathcal W}_{class}}{\partial h}$,
at $J_x \beta = 0.8$, for the ferromagnetic case 
(with $J_y/J_x = 1/3$, $J_z/J_x = -2/3$, $D/J_x= -0.6$) 
and the antiferromagnetic case 
(with $J_y/J_x = 1/3$, $J_z/J_x = 2/3$, $D/J_x=-0.6$). 
For these same cases, in   Fig. \ref{fig_11} we compare the 
classic magnetic susceptibility per  site at $h/J_x=0.35$.
  
Since we are working with unitary classical spins ($|\vec{s}_i| =1$), we
define $\langle \cos \theta_i \rangle \equiv \langle \vec{s}_i \cdot
\vec{s}_{i+1} \rangle$, where $\theta_i$ is the angle between
the $i$-th and $(i+1)$-th    spins in the chain. 
In Fig. \ref{fig_12}a we plot the  function 
$\langle \cos \theta_i \rangle \equiv 
\left( \frac{\partial}{\partial J_x} +
\frac{\partial}{\partial J_y} +
\frac{\partial}{\partial J_z} \right) {\mathcal W}_{class}$ 
for the antiferromagnetic case
(with $J_y/J_x = 1/3$, $J_z/J_x = 2/3$, $D/J_x= -0.6$), whereas 
 in Fig. \ref{fig_12}b 
we plot $\langle \cos \theta_i \rangle$ for the  ferromagnetic case 
(with $J_y/J_x = 1/3$, $J_z/J_x = -2/3$, $D/J_x= -0.6$), 
 both at $h/J_x= 0.35$. From Fig. \ref{fig_12}b  we see that 
 at $J_x \beta = 0.9$,  on the average, neighboring spins  are
orthogonal to each other. For these cases, we also plot in Fig.
\ref{fig_13} the correlation function 
$\langle s^z_i s^z_{i+1}\rangle$.

Within the range of the independent variable ($h/J_x$) shown in Fig.
\ref{fig_10}, the HTE of the magnetization $M$ is almost 
equal to its exact solution. This range has been determined
so that the HTEs of leading orders $\beta^{11}$ and $\beta^{12}$
differs by $0.1\%$  therein. A similar
thought guided the determination of the ranges of $J_x\beta$ in Figs.
\ref{fig_11} and \ref{fig_13}, regarding the magnetic susceptibility
$\chi$ and correlation function $\langle s^z_i s^z_{i+1}\rangle$,
respectively.

\section{Conclusions}

The method developed in Ref. \cite{chain_m} can be equally applied to
both quantum and classical chains with first-neighbor interactions, 
spatial periodic boundary conditions and translational invariance. 
In Ref. \cite{brief}  we calculated the high temperature
 expansion (HTE)  of the Helmholtz free  energy (HFE), 
up to order $\beta^5$,  of the quantum spin-$S$ $XYZ$ 
chain, with a single-ion anisotropy term
and in the presence of an external magnetic 
field.  From this result, we obtained the HTE of the 
classical version of the model by taking the limit $S\rightarrow \infty$.
In the present paper, we apply the method  of Ref. \cite{chain_m} directly
to the classical $XYZ$  (also with  a single-ion anisotropy term and in the
presence of an external magnetic  field) thus simplifying 
enormously algebraic calculations. By this way, we are able to calculate the 
$\beta$-expansion of its HFE up  to order $\beta^{12}$.  
Each coefficient of $\beta^n$ in the expansion is exact ($n = -1, 0,
1, 2, \cdots, 12$). Having a higher order in $\beta$ allowed extending
the knowledge of the thermodynamics of the classical $XYZ$ chain up to
$kT/J_x \sim 0.5$, which might be considered
as an intermediate region of temperature. 
Joyce's  exact result \cite{joyce} of the classical $XXZ$ chain, 
with no single-ion anisotropy term and no external magnetic field ($D=0, h=0$),
permits knowing   the behaviour of the specific heat per site at very
low  temperatures.  This low-temperature information can 
be combined with  the high-temperature information by 
the two-point Pad\'e approximants (PA)\cite{granroth}. The best PA
gives a very good description of the classical specific heat per site
in the whole interval of temperature.

In the high temperature region ($J_x \beta \ll 1$), the quantum
spin-$S$ $XYZ$ chain ($S= 1/2, 1, 3/2, 2, \cdots$) is well 
described by the classical model. Certainly, the best application of the 
HFE of the classical $XYZ$ chain is to support the  study  the
thermodynamical  properties of its quantum models for finite spin-$S$
at lower temperatures. For $S\ge 2$ we have very few numerical analysis
of the thermodynamical functions of these quantum models due to the 
large number of degrees of freedom to be handled.  By using   the Monte Carlo
numerical calculation done by  Yamamoto\cite{yamamoto96}, for 
a   spin-2 antiferromagnetic   chain  with 96 sites,  we 
showed that its specific heat and magnetic susceptibility per site
can be approximated by their respective classical 
results,  up to  $kT/J_x \approx 0.8$, within a precision of $2.5\%$. 
This gives us hope  that the thermodynamics of the TMCC ($S=5/2$) can be 
approximated by the HFE presented in this paper, up to
this temperature, within a precision higher than $2.5\%$.

Finally, from the HTE for the HFE of the classical $XYZ$ 
chain, which can be found in Eq.(\ref{B1}) of Appendix \ref{Apend_B},
we verified  that in the absence of an external magnetic field the
quantum ferromagnetic and antiferromagnetic chains have the same 
classical limit. As a consequence of this essentially classical result,
some thermodynamical functions like: mean energy, 
$\langle (s_i^z)^2)\rangle$, entropy and the correlation functions 
between  first-neighbor spin components  have defined
parity, at $h=0$, under the  parameter transformation
$(J_x, J_y, J_z) \rightarrow (-J_x, -J_y, -J_z)$. This fact permits 
studying  the reaction of each type of chain (ferromagnetic and 
antiferromagnetic) to the presence of an external magnetic field.


\begin{acknowledgments}
The authors are in debt to CNPq for partial financial support.
S.M. de S.  and O.R. thank FAPEMIG and M.T.T. thanks FAPERJ for partial
financial support.
\end{acknowledgments}

\appendix

\setcounter         {equation}{0}
\def\theequation{A.\arabic{equation}}

\section{Useful integrals } \label{Apend_A}

The normalized traces of products of operators are expressed in terms
of surface integrals over unitary spheres, each of which represents the state space
of a chain site. For example,

\begin{equation}
\langle {\textbf H}_{1,2} \rangle \equiv 
\int_{S1}\!\! d\tilde \Omega_1 \int_{S2}\!\! d\tilde \Omega_2 
\ {\cal H}(\theta_1, \phi_1, \theta_2, \phi_2),
\end{equation}

\noindent where the normalized solid angles $d \tilde \Omega_i$ 
are defined\cite{takahashi} as

\begin{equation}
d \tilde \Omega_i \equiv \frac{\sin \theta_i}{ 4 \pi} \, d \theta_i \,
d \phi_i,
\end{equation}

\noindent and the function ${\cal H}$ has been defined in Eq.(\ref{h_angles}). 
We have $\theta_i \in [0, \pi]$ and $\phi_i \in [0,2\pi]$. Another example is

\begin{eqnarray}
\langle {\textbf H}_{1,2}^2 {\textbf H}_{2,3}  \rangle &\equiv& 
\int_{S1}\!\! d\tilde \Omega_1 \int_{S2}\!\! d\tilde \Omega_2 \int_{S3}\!\!  d\tilde \Omega_3
\ {\cal H}(\theta_1, \phi_1, \theta_2, \phi_2)^2 \times \nonumber \\
& & \times \ {\cal H}(\theta_2, \phi_2,\theta_3, \phi_3).
\end{eqnarray}

\noindent The generalization of these formulas is straightforward. Because of the 
structure of the Hamiltonian (\ref{1}) and the expression of spin
operators (\ref{2}), the integrals related to the 
$i$-th site which contribute to $H_{1,m}^{(n)}$ (see Eq. (\ref{6})), 
turn out to be either 

\begin{eqnarray} \label{A1}
		%
		%
	&& I_{\theta}^{(m,n)} \equiv \int_{0}^{ \pi} d \theta\  (\sin \theta)^{m} (\cos \theta)^{n} \nonumber \\
		%
		%
	&=&\frac{1}{4} (1+(-1)^m) (1+(-1)^n) 
	\frac{\Gamma(\frac{m+1}{2})\Gamma(\frac{n+1}{2})}{\Gamma(\frac{m+n+2}{2})}  \nonumber\\
		%
		%
	& & + \frac{1}{2} (1-(-1)^m) \frac{(1+(-1)^n)}{n+1}
	\frac{\Gamma(\frac{n+3}{2})\Gamma(\frac{m+1}{2})}{\Gamma(\frac{m+n+2}{2})},  
	                     \nonumber \\
\end{eqnarray}

\noindent or

\begin{eqnarray}\label{A2}
		%
		%
	&& I_{\phi}^{(m,n)} \equiv \int_{0}^{ 2 \pi}\, d \phi \ (\sin \phi)^{m}  (\cos \phi)^{n}  \nonumber \\
		%
		%
	 &= & \frac{1}{2} (1+(-1)^m) (1+(-1)^n)\frac{\Gamma(\frac{m+1}{2}) \Gamma(\frac{n+1}{2})}  
	       {\Gamma(\frac{m+n+2}{2})},  \nonumber\\
\end{eqnarray}

\noindent where  $n, m = 0,1, 2, \dots$ and $\Gamma$ is the gamma function.


\setcounter         {equation}{0}
\def\theequation{B.\arabic{equation}}

\section{The $\beta$-expansion of the  Helmholtz Free Energy in the 
absence of a magnetic field} \label{Apend_B}

The high temperature expansion of the HFE of the classical $XYZ$  ($S\rightarrow \infty$)
in the  absence of an external magnetic field (Hamiltonian (\ref{1}) with
$h=0$), up to order $\beta^6$, is

\begin{widetext}  

\begin{eqnarray} \label{B1}
{\mathcal W}_{class}^{h=0}& =& 
- \frac{ \mathrm{ln}(2S)}{\beta}
 + \frac {\mathrm{D}}{3}  + ( -   \frac {\mathit{J_z}^{2}}{18}
 -  \frac {\mathit{J_y}^{2}}{18}  -  \frac {\mathit{J_x}^{2}}{18}
 -  \frac {2\,\mathrm{D}^{2}}{45} )\,\beta      \nonumber  \\
&+ & ( -  \frac {2\,\mathit{J_y}^{2}\,\mathrm{D}}{135}
 -  \frac {2\,\mathit{J_x}^{2}\,\mathrm{D}}{135}
  +  \frac {8\,\mathrm{D}^{3}}{2835}  + 
 \frac {4\,\mathit{J_z}^{2}\,\mathrm{D}}{135} )\,\beta ^{2} 
 + ( \frac {4\,\mathrm{D}^{4}}{14175} 
 -  \frac {32\,\mathit{J_z}^{2}\,\mathrm{D}^{2}}{4725}  \nonumber \\
& + &  \frac {2\,\mathit{J_y}^{2}\,\mathrm{D}^{2}}{4725}
-  \frac {7\,\mathit{J_x}^{4}}{2700}
 - {\displaystyle \frac {7\,\mathit{J_z}^{4}}{2700}} 
  +  \frac {\mathit{J_x}^{2}\,\mathit{J_z}^{2}}{225}
  +  \frac {2\,\mathit{J_x}^{2}\,\mathrm{D}^{2}}{4725} 
   +  \frac {\mathit{J_x}^{2}\,\mathit{J_y}^{2}}{225} 
    - \frac {7\,\mathit{J_y}^{4}}{2700}   \nonumber \\
& + &   \frac {\mathit{J_y}^{2}\,\mathit{J_z}^{2}}{225} )\beta ^{3}
 + (\frac {16\,\mathit{J_y}^{2}\,\mathrm{D}^{3}}{42525} 
 -  \frac {2\,\mathit{J_y}^{4}\,\mathrm{D}}{2835}
 -  \frac {32\,\mathrm{D}^{5}}{467775}  
 +  \frac {4\,\mathit{J_z}^{4}\,\mathrm{D}}{2835}
  -  \frac {2\,\mathit{J_x}^{4}\,\mathrm{D}}{2835}  \nonumber  \\
&+&   \frac {16\,\mathit{J_x}^{2}\,\mathrm{D}^{3}}{42525}  
+  \frac {8\,\mathit{J_x}^{2}\,\mathit{J_y}^{2}\,\mathrm{D}}{2835}
-  \frac {4\,\mathit{J_y}^{2}\,\mathit{J_z}^{2}\,\mathrm{D}}{2835}
 + \frac {16\,\mathit{J_z}^{2}\,\mathrm{D}^{3}}{42525}  
 -  \frac {4\,\mathit{J_x}^{2}\,\mathit{J_z}^{2}\,\mathrm{D}}{2835} )\beta ^{4}
                               \nonumber \\
 &+&  ( -  \frac {736\,\mathrm{D}^{6}}{1915538625} 
 -  \frac {107\,\mathit{J_x}^{6}}{2679075}  
 -  \frac {107\,\mathit{J_y}^{6}}{2679075}  
 -  \frac {107\,\mathit{J_z}^{6}}{2679075} 
  -  \frac {844\,\mathit{J_z}^{2}\,\mathit{J_x}^{2}\,\mathit{J_y}^{2}}{893025} 
                                \nonumber \\
& + & \frac {106\,\mathit{J_z}^{4}\,\mathit{J_x}^{2}}{893025}  
+  \frac {106\,\mathit{J_z}^{4}\,\mathit{J_y}^{2}}{893025} 
+  \frac {106\,\mathit{J_x}^{4}\,\mathit{J_y}^{2}}{893025} 
 +  \frac {106\,\mathit{J_x}^{2}\,\mathit{J_y}^{4}}{893025}  \nonumber \\
& +&  \frac {106\,\mathit{J_z}^{2}\,\mathit{J_y}^{4}}{893025}  
+ \frac {106\,\mathit{J_z}^{2}\,\mathit{J_x}^{4}}{893025}
+  \frac {8768\,\mathit{J_z}^{2}\,\mathrm{D}^{4}}{49116375}  
+  \frac {106\,\mathrm{D}^{2}\,\mathit{J_x}^{4}}{1488375} 
 +  \frac {106\,\mathrm{D}^{2}\,\mathit{J_y}^{4}}{1488375}   \nonumber \\
& - &  \frac {16\,\mathit{J_z}^{4}\,\mathrm{D}^{2}}{297675}
-  \frac {2008\,\mathrm{D}^{4}\,\mathit{J_y}^{2}}{49116375}  
-  \frac {2008\,\mathrm{D}^{4}\,\mathit{J_x}^{2}}{49116375}  
-  \frac {232\,\mathit{J_z}^{2}\,\mathrm{D}^{2}\,\mathit{J_x}^{2}}{1488375}  
                                            \nonumber \\
& - & \frac {232\,\mathit{J_z}^{2}\,\mathrm{D}^{2}\,\mathit{J_y}^{2}}{1488375} 
 +  \frac {584\,\mathrm{D}^{2}\,\mathit{J_x}^{2}\,\mathit{J_y}^{2}}{1488375} )\beta ^{5} 
 + ( \frac {2944\,\mathrm{D}^{7}}{1915538625} 
 -  \frac {736\,\mathrm{D}^{5}\,\mathit{J_x}^{2}}{70945875}   \nonumber \\
&- &  \frac {25024\,\mathit{J_z}^{2}\,\mathrm{D}^{5}}{638512875}  
-  \frac {736\,\mathrm{D}^{5}\,\mathit{J_y}^{2}}{70945875}
 -  \frac {8912\,\mathit{J_z}^{4}\,\mathrm{D}^{3}}{49116375}  
 +  \frac {52\,\mathrm{D}\,\mathit{J_y}^{6}}{1913625}  
 + \frac {52\,\mathrm{D}\,\mathit{J_x}^{6}}{1913625}  \nonumber \\
& + &  \frac {904\,\mathrm{D}^{3}\,\mathit{J_x}^{4}}{49116375}  
+  \frac {904\,\mathrm{D}^{3}\,\mathit{J_y}^{4}}{49116375} 
 -  \frac {104\,\mathit{J_z}^{6}\,\mathrm{D}}{1913625}  
 +  \frac {6976\,\mathit{J_z}^{2}\,\mathrm{D}^{3}\,\mathit{J_x}^{2}}{49116375}  
                     \nonumber\\
&+ &  \frac {6976\,\mathit{J_z}^{2}\,\mathrm{D}^{3}\,\mathit{J_y}^{2}}{49116375} 
-  \frac {4736\,\mathrm{D}^{3}\,\mathit{J_x}^{2}\,\mathit{J_y}^{2}}{49116375}  
+  \frac {52\,\mathit{J_z}^{4}\,\mathrm{D}\,\mathit{J_y}^{2}}{637875} 
 -  \frac {52\,\mathit{J_z}^{2}\,\mathrm{D}\,\mathit{J_x}^{4}}{637875} 
    \nonumber  \\
& - &  \frac {52\,\mathit{J_z}^{2}\,\mathrm{D} \,\mathit{J_y}^{4}}{637875}  
+  \frac {52\,\mathit{J_z}^{4}\,\mathrm{D}\,\mathit{J_x}^{2}}{637875} )\beta ^{6} 
+ { \mathcal O}(\beta ^{7})  .
 \end{eqnarray}

\end{widetext}

\noindent In  the   absence of  an external magnetic field ($h=0$), 
this is  an even function  of the coupling constants  
$J_x$, $J_y$ and $J_z$ of the Hamiltonian (\ref{1}). 
We have confirmed  this property up to order $\beta^{12}$.



\begin{widetext}

\begin{figure}
\includegraphics[width=10cm,height=16cm, angle=-90]{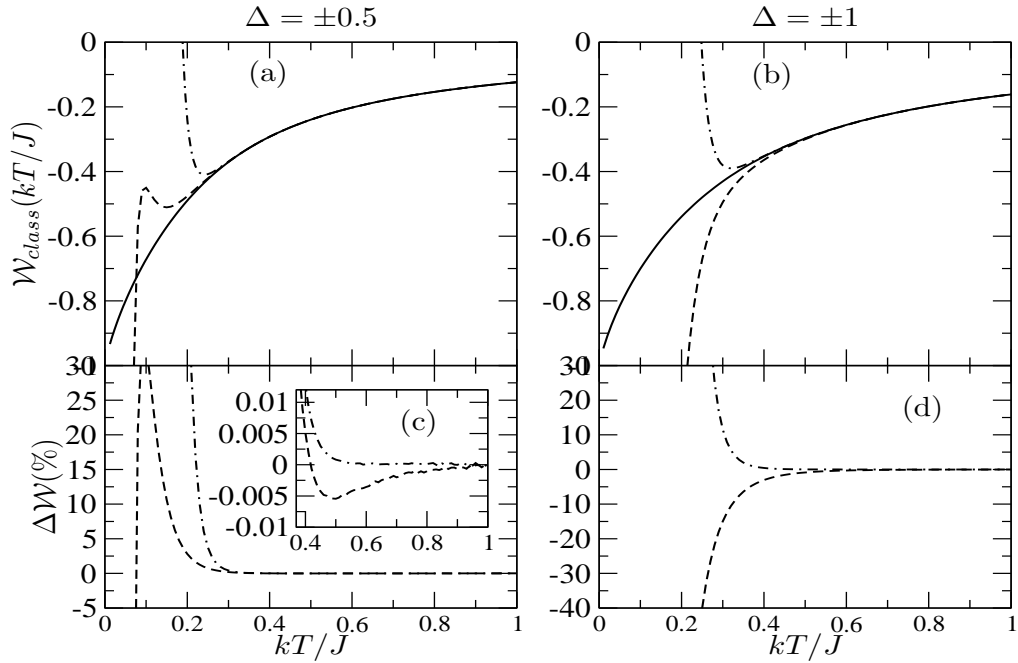}
\caption{(a) For $\Delta=\pm 0.5$, 
comparison of   the numerical solution of Joyce's
 expression (solid line) and our HTEs up
to order $\beta^5$ (dashed line) and order $\beta^{12}$ (dashed-dotted
line), for the HFE of the classical $XXZ$ chain (${\mathcal
W}_{class}$), as a function of $kT/J$.
In (c),   the  relative percental differences 
($\Delta{\cal W}(\%)$), with respect to Joyce's exact
solution, of the $\beta^5$ (dashed line) and the $\beta^{12}$
(dashed-dotted line) expansions, also as functions of $kT/J$.
In (b) and (d) a similar comparison is performed, for $\Delta = \pm 1$.}
\label{fig_1}
\end{figure}

\begin{figure}
\includegraphics[width=10cm,height=16cm,angle=-90]{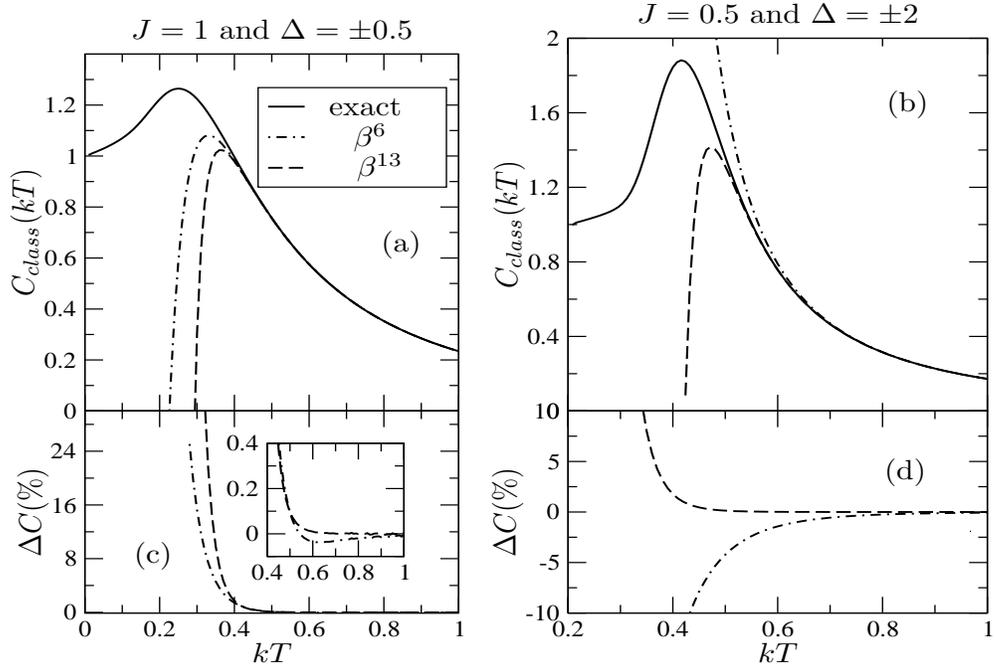}
\caption{(a) Specific heat per site $C_{class}$ of the classical
$XXZ$ chain, as a function of $kT$, $T$ being the absolute temperature,
for $J=1$ and $\Delta= \pm 0.5$. 
The upper pane shows Joyce's solution (solid line),
and  our expansions up to  $\beta^6$ (dotted-dashed line)\cite{brief} 
and up to $\beta^{13}$ (dashed  line). The lower 
pane (c)   shows the percental
differences ($\Delta C (\%)$), with respect to Joyce's solution, of the
$\beta^6$ (dotted-dashed line) and
 $\beta^{13}$ (dashed line) expansions, also as functions of $kT$. 
Panes (b) and (d) show a  similar comparison, 
for $J=0.5$ and $\Delta= \pm 2$.}
\label{fig_2}
\end{figure}

\begin{figure}
\includegraphics[width=10cm,height=16cm,angle=-90]{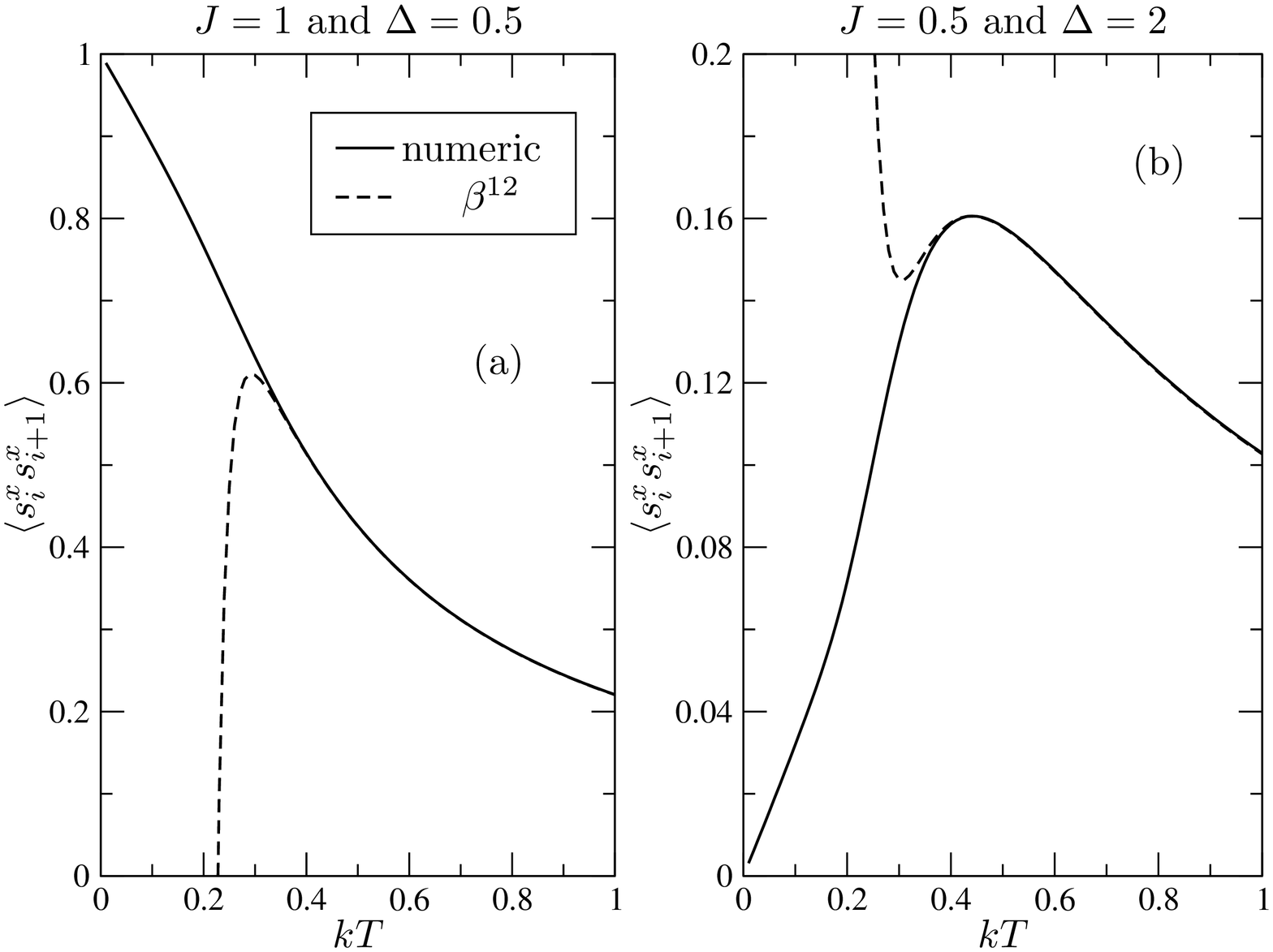}
\caption{ The correlation function  $\langle s_i^x s_{i+1}^x \rangle$ 
between the classical spin $x$-components of first neighbouring sites.
The solid line stands for the exact function, whereas the
dashed line stands for the HTE up to order $\beta^{12}$. In (a) we have $J=1$ and
$\Delta= 0.5$; in (b) we have $J=0.5$ and $\Delta= 2$. }
\label{fig_3}
\end{figure}

\begin{figure}
\includegraphics[width=10cm,height=16cm,angle=-90]{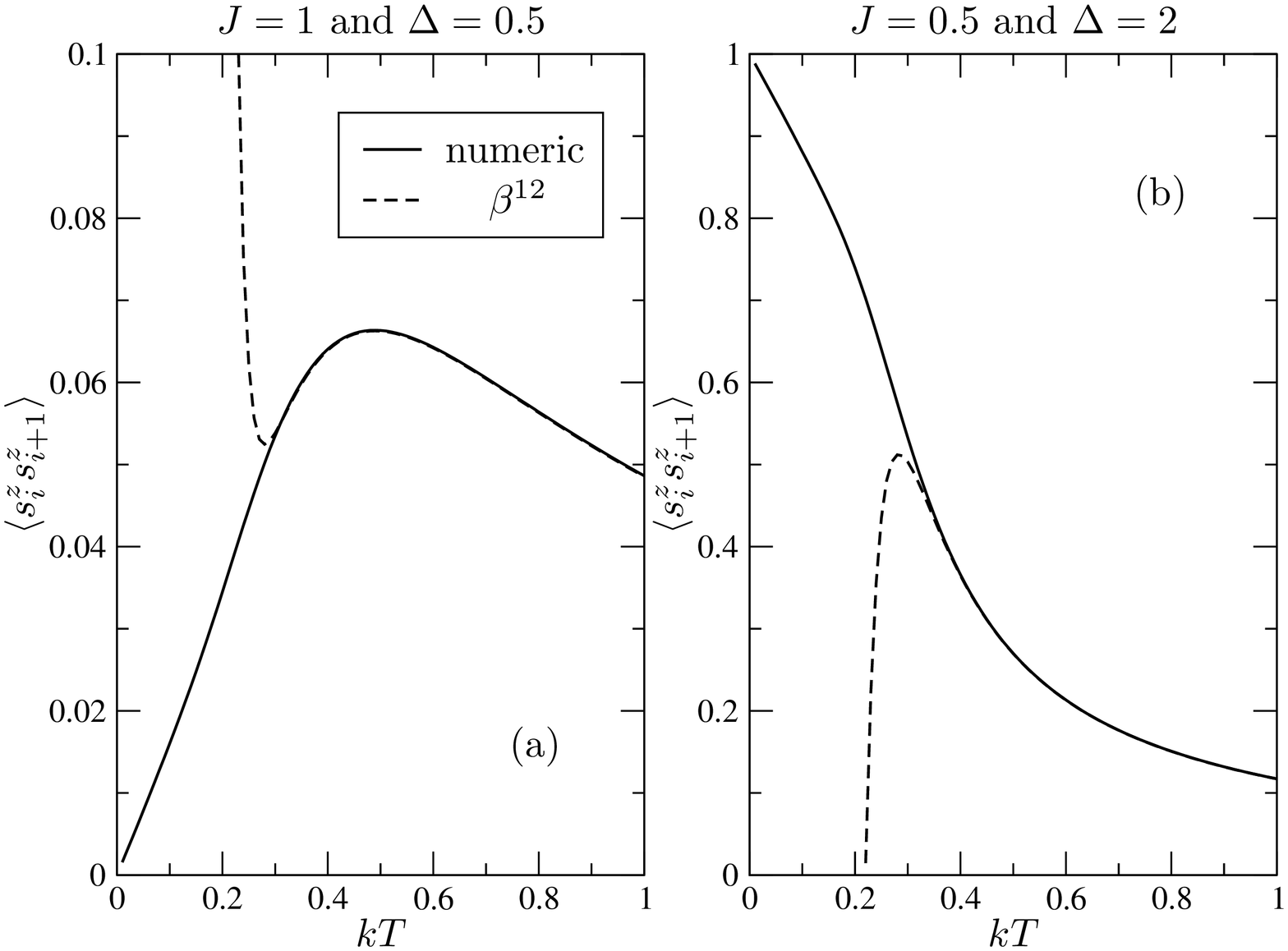}
\caption{ The correlation function  $\langle s_i^z s_{i+1}^z \rangle$ 
between the classical spin $z$-components of first neighbouring sites.
The solid line stands for the exact function, whereas the
dashed line stands for the HTE up to order $\beta^{12}$. In (a) we have $J=1$ and
$\Delta= 0.5$; in (b) we have $J=0.5$ and $\Delta= 2$.}
\label{fig_4}
\end{figure}

\begin{figure}
\includegraphics[width=10cm,height=16cm,angle=-90]{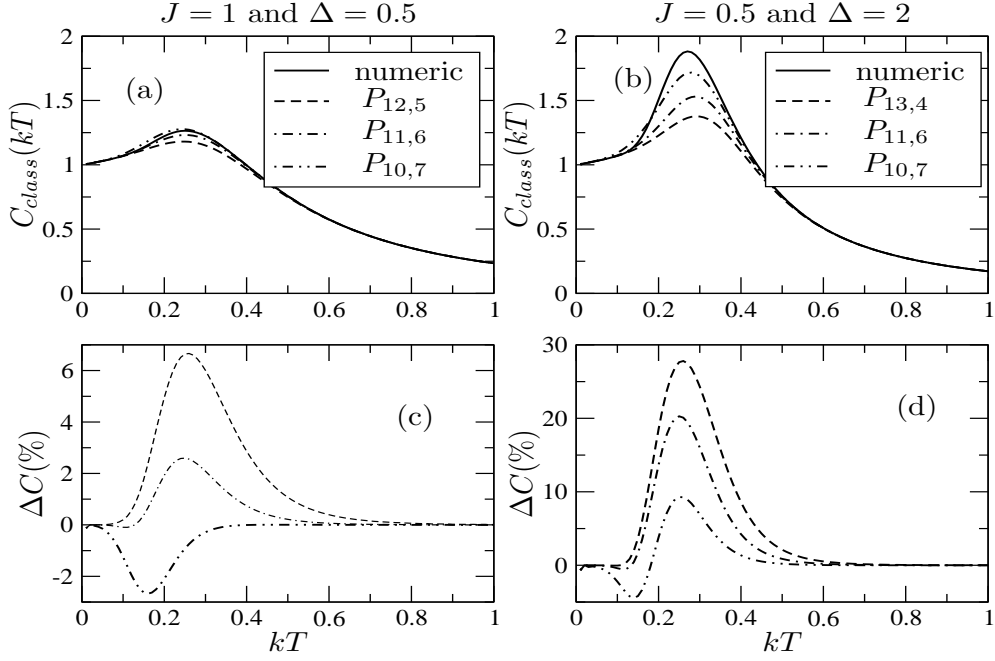}
\caption{Upper  panes (a) and (b)  compare exact
 classical specific heat per site and their
best PAs. Lower  panes (c) and (d)  show the 
percental difference of each PA with
respect to the exact result. In  (a) and (c)   we have 
$J=1$ and $\Delta =0.5$, whereas in 
(b) and (d)  we have $J=0.5$ and $\Delta=2$. }
\label{fig_5}
\end{figure}

\begin{figure}
\includegraphics[width=10cm,height=16cm,angle=-90]{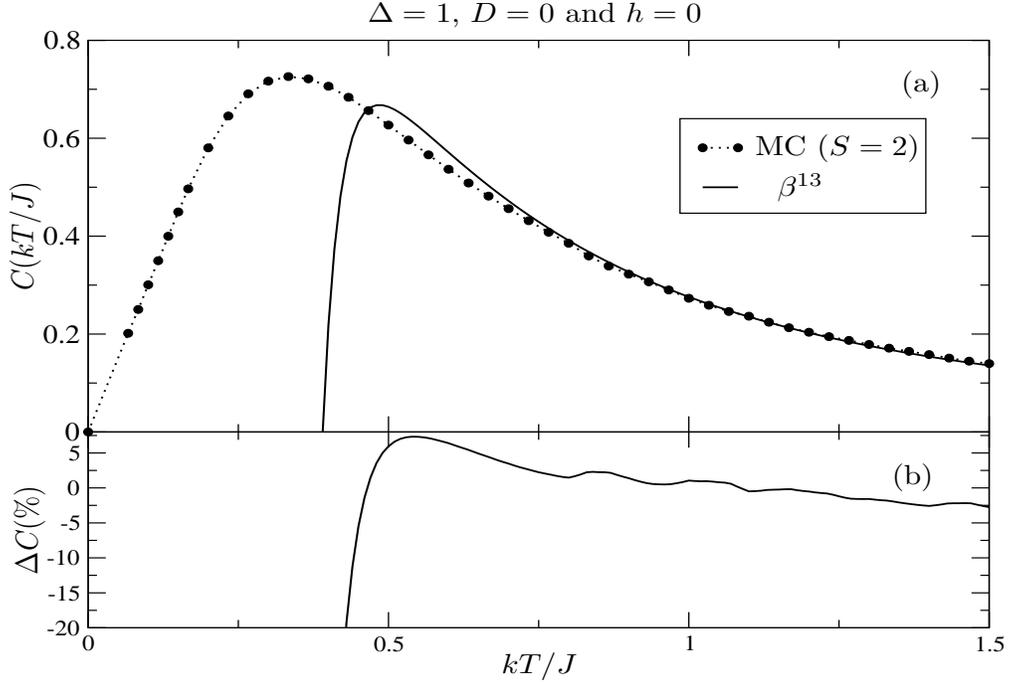}
\caption{  (a) The  dots correspond to the Monte Carlo (MC)
calculation of the specific heat per site for the  $S=2$
antiferromagnetic chain\cite{yamamoto96}, for $D=0$ and $h=0$, as a
function of  $kT$,  where $T$ is  the absolute temperature. 
The solid line corresponds  to our HTE of the classical 
specific heat up to order $\beta^{13}$.   (b) The  relative percental 
difference of the two curves.}
\label{fig_6}
\end{figure}

\begin{figure}
\includegraphics[width=10cm,height=16cm,angle=-90]{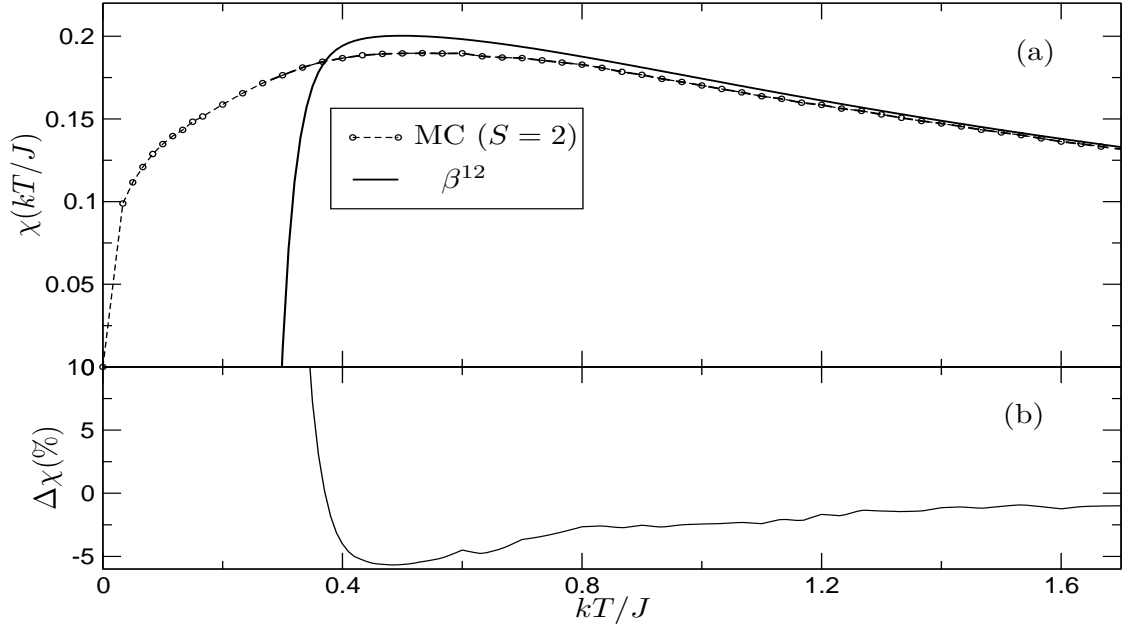}
\caption{In (a) we compare 
Yamamoto's  MC calculation of the magnetic susceptibility
 per site of the  $S=2$  antiferromagnetic
chain\cite{yamamoto96} (dots and dashed line) and 
the HTE of its classical equivalent,  up to order $\beta^{12}$ (solid
line).  In (b), we have the percental difference of the
 two functions plotted in  (a). We have $\Delta=1$, $D=0$ and $h=0$.
}
\label{fig_7}
\end{figure}

\begin{figure}
\includegraphics[width=10cm,height=16cm,angle=-90]{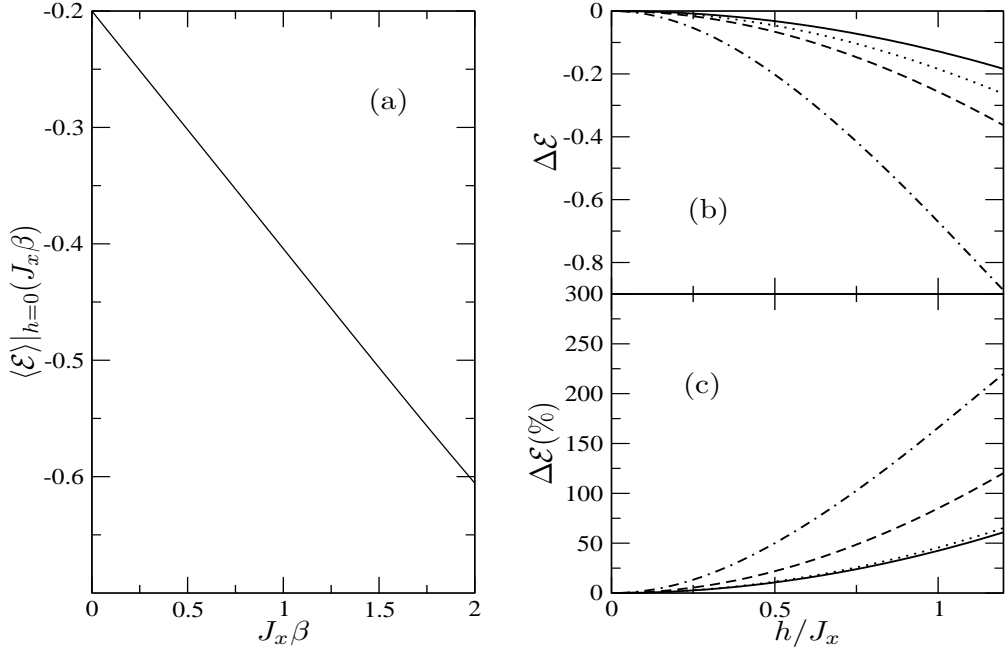}
\caption{ (a) The mean energy per site 
$\langle\varepsilon\rangle$ of  ferromagnetic 
($J_z/J_x = -2/3$)  and antiferromagnetic
($J_z/J_x = 2/3$)  
media in the absence of an external magnetic 
field.  (b) The  difference $\Delta\varepsilon$  
of the mean energy per site at finite 
$h$ and its value at $h=0$  versus $h/J_x$ . We 
plot the antiferromagnetic  case 
for  $J_x \beta = 0.5$ (solid line) and
 $J_x \beta= 1$ (dotted line); 
and also  the ferromagnetic case 
for $J_x \beta = 0.5$ (dashed line) and 
$J_x \beta= 1$ (dotted-dashed line). In (c) we have 
the correspondent percental differences 
$\Delta\varepsilon (\%)$with respect to the $h=0$ case;
the same graphical convention for lines is used. In all figures we have
$J_y/J_x = 1/3$ and $D/J_x= -0.6$.}
\label{fig_8}
\end{figure}

\begin{figure}
\includegraphics[width=10cm,height=16cm,angle=-90]{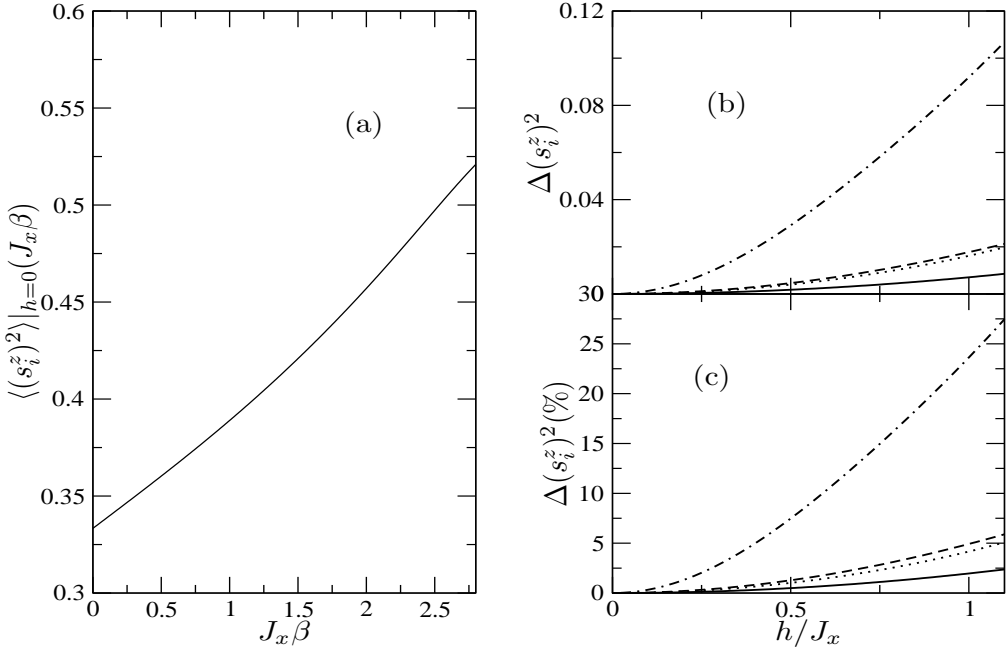}
\caption{ (a) The mean value of  the squared $z$-component
 of spin per site  $\langle (s_i^z)^2 \rangle$ 
for  ferromagnetic  ($J_z/J_x = -2/3$) and 
antiferromagnetic ($J_z/J_x = 2/3$)  media  is
the same, in the absence of an external magnetic field. (b) The
difference of $\langle (s_i^z)^2 \rangle$ at finite $h$ and its value
at $h=0$ versus $h/J_x$ . We plot the antiferromagnetic case at $J_x
\beta = 0.5$ (solid line) and $J_x \beta= 1$ (dotted line); 
and the ferromagnetic  case at $J_x \beta = 0.5$ (dashed line) 
and $J_x \beta= 1$ (dotted-dashed line).  {Fig. (c) shows the corresponding
 percental differences with respect to the $h=0$ case; 
the same graphical convention for lines is used.} In all figures 
we have $J_y/J_x = 1/3$ and   $D/J_x= -0.6$. }
\label{fig_9}
\end{figure}

\begin{figure}
\includegraphics[width=10cm,height=16cm,angle=-90]{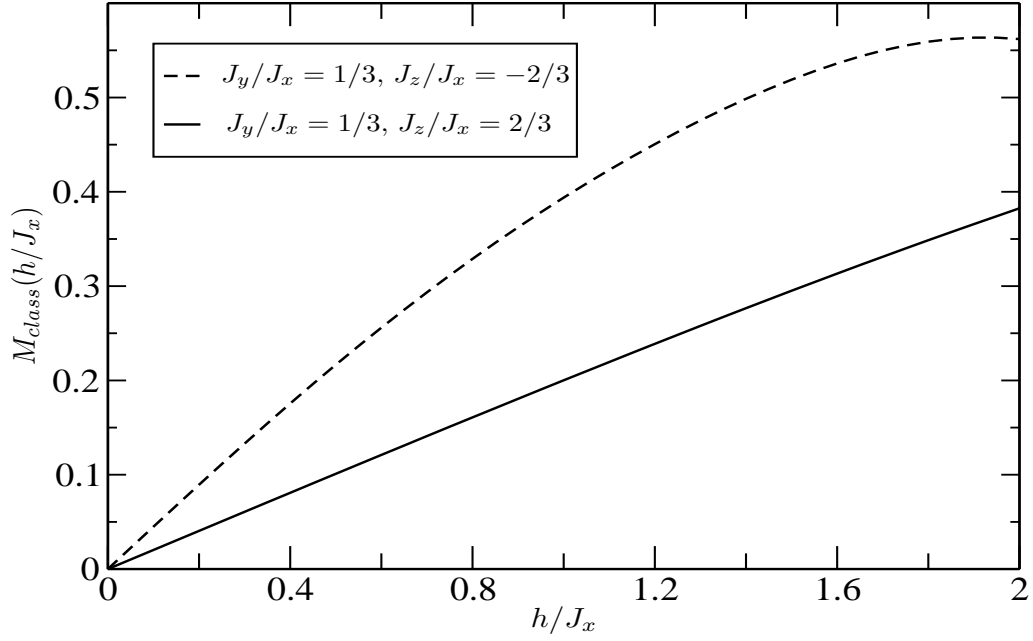}
\caption{  The classical magnetization per site  $M_{class}$  
as a function of $h/J_x$ with  $J_y/J_x = 1/3$ and 
$D/J_x= -0.6$ at $J_x \beta= 0.8$, for the
ferromagnetic ($J_z/J_x = -2/3$, dashed line) and 
antiferromagnetic   ($J_z/J_x =2/3$, solid line)  cases.}
\label{fig_10}
\end{figure}

\begin{figure}
\includegraphics[width=10cm,height=16cm,angle=-90]{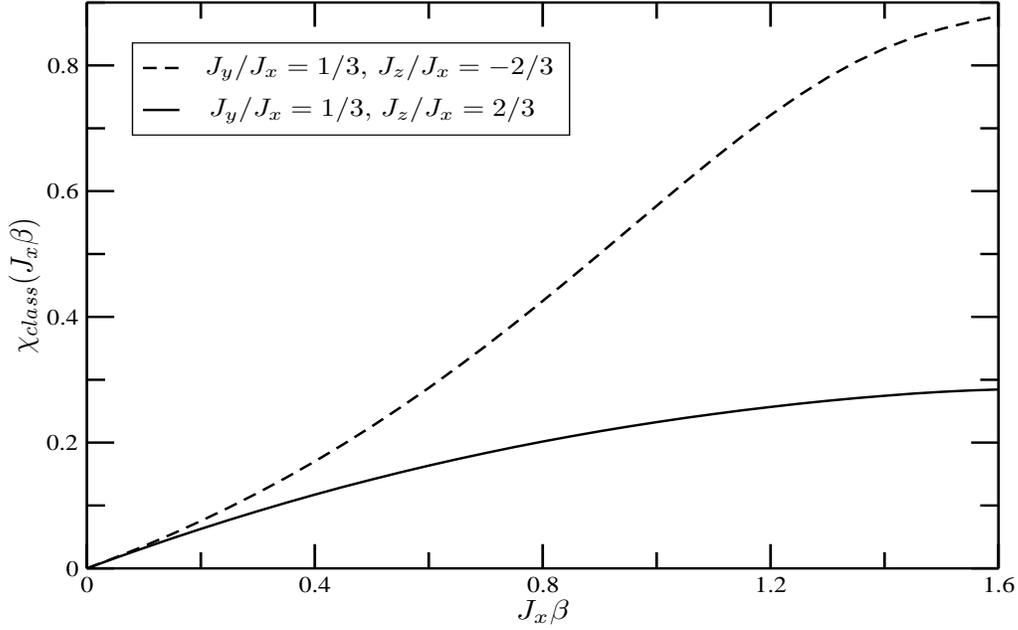}
\caption{ The classical  magnetic susceptibility per site 
$\chi_{class}$ as a function of   $J_x \beta$ with $J_y/J_x = 1/3$ and
 $D/J_x = -0.6$ at $h/J_x = 0.35$, for  the ferromagnetic 
 ($J_z/J_x = -2/3$, dashed line) and antiferromagnetic 
($J_z/J_x =2/3$, solid line) cases. }
\label{fig_11}
\end{figure}

\begin{figure}
\includegraphics[width=10cm,height=16cm,angle=-90]{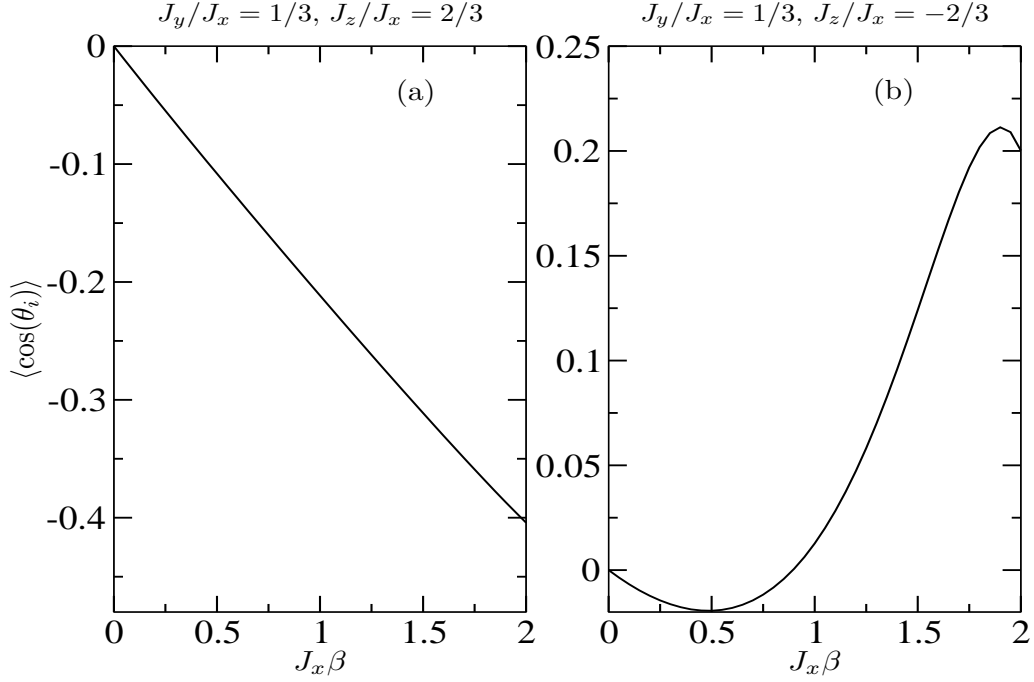}
\caption{ The function $\langle\cos \theta_i\rangle$ as a function of
$J_x \beta$, for $i= 1, 2, \cdots, N$   with  $J_y/J_x = 1/3$ and $D/J_x =
-0.6$ at $h/J_x =0.35$. Fig. (a) shows the  antiferromagnetic   case
($J_z/J_x = 2/3$) and Fig. (b) the  ferromagnetic  case 
($J_z/J_x = - 2/3$).}
\label{fig_12}
\end{figure}

\begin{figure}
\includegraphics[width=10cm,height=16cm,angle=-90]{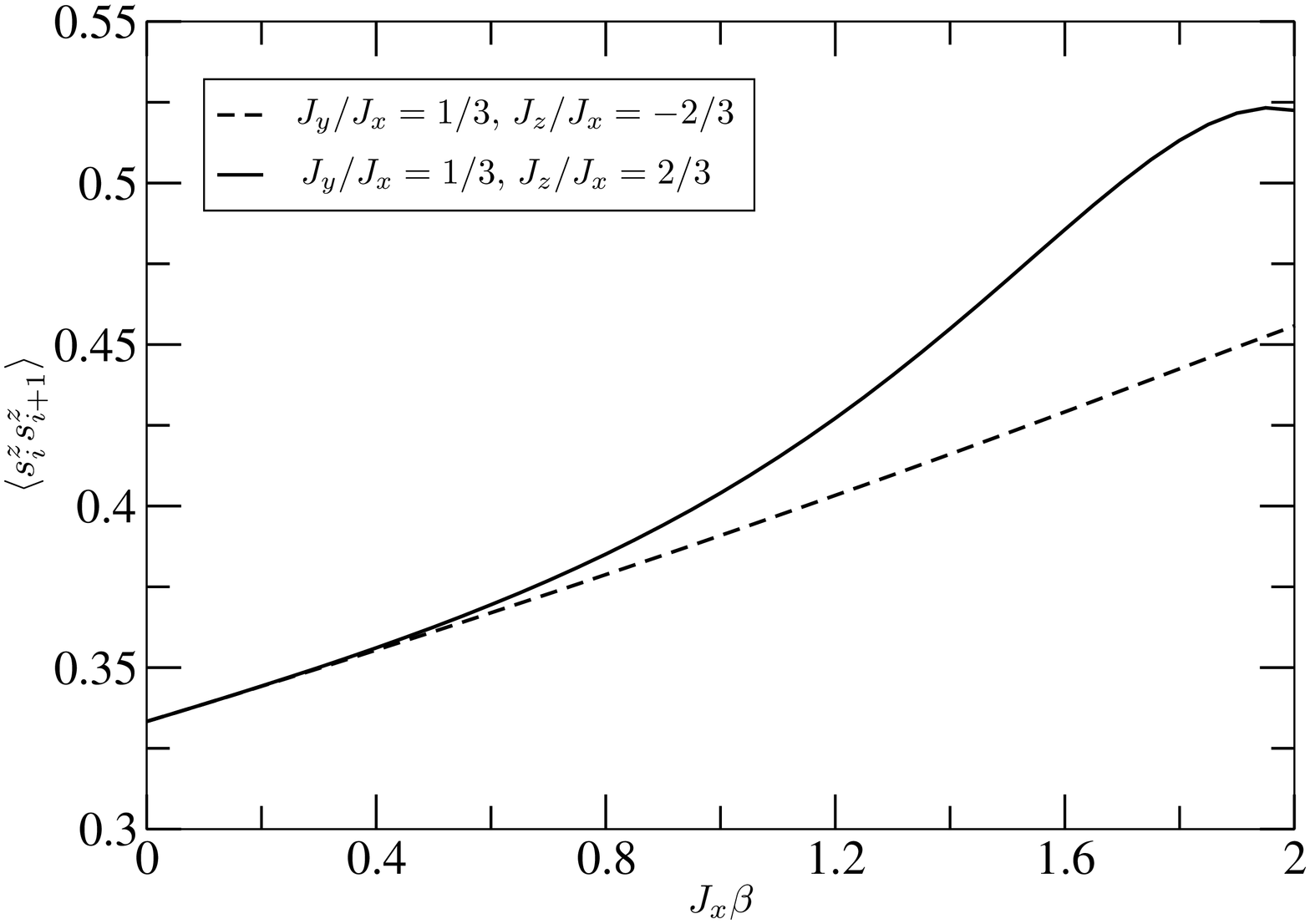}
\caption{The correlation function $\langle s^z_i s^z_{i+1}\rangle$
between the spin $z$-components of first neighbors as a
function of $J_x \beta$. The dashed and solid lines describe the
ferromagnetic case ($J_z/J_x = -2/3$) and antiferromagnetic case
($J_z/J_x = 2/3$), respectively. We take $J_y/J_x = 1/3$ and $D/J_x=
-0.6$ at $h/J_x = 0.35$. }
\label{fig_13}
\end{figure}


\end{widetext}


\begin{thebibliography}{}

\bibitem{fisher} M.E. Fisher, Am. J. of Phys. {\bf 32},  343 (1964).

\bibitem{joyce} G. S. Joyce, Phys. Rev. Lett. {\bf 19}, 581 (1967).

\bibitem{sine-gordon}  H.J. Mikeska, J. Phys.: Solid St. Phys. {\bf 13}, 2913 (1980), 
and references therein.

\bibitem{delica} T. Delica, and H. Leschke, Physica A {\bf 168}, 768 (1990).

\bibitem{brief}   O. Rojas,  S. M. de Souza,  E. V. Corr\^ea Silva, and 
M. T. Thomaz, Phys. Rev. B {\bf 72}, 172414  (2005).

\bibitem{chain_m} O. Rojas, S. M. de Souza and M. T. Thomaz,
 J. Math. Phys. {\textbf 43}, 1390 (2002).
 
\bibitem{prb67_03}  O. Rojas,   E. V. Corr\^ea Silva,  W.A. Moura-Melo,  S. M. de Souza, and 
M. T. Thomaz, Phys. Rev. B {\bf 67}, 115128 (2003).

\bibitem{ejpB} O. Rojas,  S. M. de Souza, E. V. Corr\^ea Silva and 
M. T. Thomaz,  Eur. Jour. of Phys. {\bf B47}, 165 (2005).

\bibitem{outros} A. B\"uller, U. L\"ow, G. S. Uhrig, Phys. Rev. {\bf B64}, 024428, (2001); 
B. Bernu and G. Misguich, Phys. Rev. {\bf B63}, 134409 (2001).

\bibitem{pr_antigo} G. A. Baker Jr., G. S. Rushbrooke, H. E. Gilbert,
 Phys. Rev. {\bf 135},  A1272 (1964).

\bibitem {granroth} G. E. Granroth, M. W. Meisel, M. Chaparala, 
Th. Jolic\oe ur, B. H. Ward and D. R. Taham, Phys. Rev. Lett. 
{\bf 77}, 1616 (1996).

\bibitem{birgeneau} R. J. Birgeneau, R. Dingle, M. T. Hutchings, 
G. Shirane and S. L. Holt, Phys. Rev. Lett. {\bf 26}, 718 (1971).

\bibitem{hutchings} M. T. Hutchings, G. Shirane, R. J. Birgeneau 
and S. L. Holt, Phys. Rev. B {\bf 5}, 1999 (1972).

\bibitem{yamamoto96} S. Yamamoto, Phys. Rev. B {\bf 53}, 3364 (1996-II); 
Phys. Lett. A {\bf 213}, 102 (1996).

\bibitem{jensen} H.J. Jensen, O.G. Mouritsen, H.C. Fogedby, P. Hedeg\aa rd
and A. Svane, Phys. Rev. B  {\bf 32}, 3240 (1985).

\bibitem{takahashi}{M. Takahashi. {\it Thermodynamics of One-dimensional Solvable
Models}. Cambridge, Cambridge University Press, 1999.}

\end{thebibliography}
\end{document}